\documentclass[]{elsart}
\usepackage{amsmath,amsfonts,amssymb,graphicx,subfigure,overpic,multirow,url}

\pdfoutput=1

\begin{document}

\begin{frontmatter}

\title{Towards constructing one-bit binary adder in excitable chemical medium}
\author{Ben De Lacy Costello, Andy Adamatzky, Ishrat Jahan, Liang Zhang}
\address{University of the West of England, Bristol, United Kingdom}

\date{\today}

\begin{abstract}

Light-sensitive modification (ruthenium catalysed) of the Belousov-Zhabotinsky medium exhibits various regimes of excitability depending on the levels of illumination. For certain values of illumination the medium switches to a sub-excitable mode. An asymmetric perturbation of the medium leads to formation of a travelling localized excitation, a wave-fragment which moves along a predetermined trajectory, ideally preserving its shape and velocity. To implement collision-based computing with such wave-fragments we represent values of Boolean variables in presence/absence of a wave-fragment at specific sites of medium. When two wave-fragments collide they either annihilate, or form new wave-fragments. The trajectories of the wave-fragments after the collision represent a result of the computation, e.g. a simple logical gate. Wave-fragments in the sub-excitable medium are famously difficult to control. Therefore, we adopted a hybrid procedure in order to construct collision-based logical gates: we used channels, defined by lower levels illumination to subtly tune the shape of a propagating wave-fragment and allow the wave-fragments to collide at the junctions between channels. Using this methodology we were able to implement both in theoretical models (using the Oregonator) and in experiment two interaction-based logical gates and assemble the gates into a basic 
one-bit binary adder. We present the first ever experimental approach towards constructing arithmetical circuits in spatially-extended excitable chemical systems.

\vspace{0.5cm}

\end{abstract}

\maketitle

\end{frontmatter}

\maketitle

\section{Introduction}

In 1994 the Showalter Laboratory presented the first ever experimental implementation of logical gates in the 
spatially-extended Belousov-Zhabotinsky (BZ) system~\cite{toth1994, toth_showalter} based on the space-time interaction of 
travelling excitation waves.  The logical gates were based on the geometrical configuration of channels in which the 
excitation waves propagated. The ratio between the channel diameter and the critical nucleation radii of the excitable 
media allowed various logical schemes to be realised. The original findings by Showalter and colleagues led to several 
innovative designs of computational devices, based on geometrically-constrained excitable substrates.

Designs incorporating assemblies of channels for excitation wave propagation were used to implement logical gates for Boolean and multiple-valued logic~\cite{sielewiesiuk_2001, motoike_2003, gorecki_2009, yoshikawa_2009}, many-input logical gates~\cite{gorecki_2006, gorecki_2006a}, counters~\cite{gorecki_2003}, coincidence detectors~\cite{gorecka_2003}, and 
detectors of direction and distance~\cite{gorecki_2005, yoshikawa_2009a}.
All these chemical computing devices were realised in geometrically-constrained media where 
excitation waves propagate along defined catalyst loaded channels or tubes filled 
with the BZ reagents. The waves perform computation by interacting at the junctions between the channels. 

In all papers cited above an excitable BZ medium was employed. In an excitable medium any initial perturbation will lead 
to the formation of omni-directional wave-fronts of excitation, usually target or spiral waves. In BZ-based logical circuits 
movement of signal, represented by  an excitation wave-front is controlled by physically restricting medium to dedicated 
channels of propagation. In~\cite{adamatzky_2004_collision, andy_ben_BZ_collision, RITABEN, ben_gun, RDCBook} we developed an alternative approach by employing self-localised excitations (remotely resembling dissipative solitons) in the sub-excitable 
BZ medium to implement collision-based computing circuits~\cite{cbc}.

The paradigm  of collision-based computing originates from the computational universality of the Game of 
Life~\cite{berlekamp_1992}, conservative logic and the billiard-ball model~\cite{fredkin_toffoli_1982} with its cellular-automaton implementation~\cite{margolus}. A collision-based computer employs mobile self-localized excitations to represent quanta of information in active non-linear media. Information values, e.g. truth values of logical variables, are given by either the absence or presence of the localizations or by other parameters such as direction or velocity.  The localizations travel in space and collide with each other. The results of the collisions are interpreted as computation. There are no predetermined stationary wires, a trajectory of the travelling localization is a momentary wire. Almost any part of the reactor space can be used as a wire. Localizations can collide anywhere within this space. The localizations undergo transformations, form bound states, annihilate or fuse. Information values of localizations are transformed as a result of these collisions~\cite{cbc}.

To implement a collision-based scheme in a spatially-extended chemical medium we must employ travelling localised excitations.
Such localisations, or wave-fragments, emerge in a light-sensitive BZ medium when it is in a sub-excitable state~\cite{sendina_2001}. The ruthenium-catalyzed BZ medium shows a high degree of light-sensitivity. At some levels of illumination the medium behaves as a  classical excitable medium where a perturbation leads to the formation of omni-directional propagating waves of excitation. When the level of illumination exceeds a critical threshold no excitation persists. There is narrow interval of illumination parameters where the BZ medium is in a sub-excitable (weakly excitable) state between the non-excitable and excitable states. A perturbation of the sub-excitable medium leads to formation of localised travelling excitations, or wave-fragments. The wave-fragments travel along their predetermined trajectories and preserve their shapes and velocity vectors for some period of time. A behaviour of each fragment is determined by the exact level of illumination and the size of the fragment. The smaller fragments usually collapse, whereas large fragments usually expand. If the illumination level is at a critical level then appropriate sized fragments entering a sub-excitable media will preserve their size for appreciable distances/time intervals. See overviews of recent results in collision-based computing in BZ in~\cite{RDCBook, adamatzkyPrivmanIssue, adamatzkyKazIssue}. 

In the present paper we combine the geometrically-constrained approach with a collision-based paradigm to deal with the instability of excitation
wave-fragments. We allow the wave-fragments to travel in channels, defined by low levels of the medium's illumination, but collide in larger areas at the junctions between the channels. Thus the size of a wave-fragment is ``automatically'' maintained whilst travelling along the channel and enters the junction as a localised excitation wave with the ability to expand/contract dependent on the level of illumination. The localised excitation waves collide and interact in the junctions according to the principles of collision-based computing. 

The paper is structured as follows. In Sect.~\ref{modelling} we present details of the two-variable Oregonator model used in simulation of BZ-based logical gates. Functionality, light-control and polymorphism of simulated ballistic gates are discussed in Sect.~\ref{gatessection}. The Computational model of a one-bit half-adder is presented in Sect.~\ref{simulatedadder}. In Sect.~\ref{experimentaltechniques} we outline the experimental techniques used for implementing BZ-based logical gates. The results of the experimental implementation of the interaction gates and one-bit half-adder are discussed in Sect.~\ref{experimentalresults}. The drawbacks of the current approach and further designs are discussed in Sect.~\ref{discussion}.

\section{Modelling technique}
\label{modelling}

We employ the two-variable Oregonator equation~\cite{field_noyes_1974} adapted for light sensitivity as an analogue of the 
Belousov-Zhabotinsky (BZ) reaction with applied illumination~\cite{beato_engel}:

$$
  \frac{\partial u}{\partial t}  =  \frac{1}{\epsilon} (u - u^2 - (f v + \phi)\frac{u-q}{u+q}) + D_u \nabla^2 u 
$$
$$
  \frac{\partial v}{\partial t}  =  u - v 
$$

The variables $u$ and $v$ represent the local concentrations of activator, or excitatory component, and inhibitor,
or refractory component respectively. Parameter $\epsilon$ sets up a ratio of time scale for the variables $u$ and $v$, $q$ is a 
scaling parameter dependent on the rates of activation/propagation and inhibition, $f$ is a stoichiometric factor. 
Constant $\phi$ is the rate of inhibitor production. In the light-sensitive BZ $\phi$ represents the rate of inhibitor
production which is proportional to the intensity of illumination. We integrate the system using the Euler method with five-node Laplace operator, time step $\Delta t=0.005$ and grid point spacing $\Delta x= 0.25$, $\epsilon=0.022$, $f=1.4$, $q=0.002$. 
The equations effectively map the space-time dynamics of excitation in the BZ medium and have proved to be an invaluable tool 
for studying the dynamics of collisions between travelling localized excitations in our previous 
work~\cite{adamatzky_2004_collision, andy_ben_BZ_collision, RITABEN, ben_gun}.

\section{Ballistic gates}
\label{gatessection}

Ballistic gates were originally designed to implement collision-based computing primitives with propagating slime mould 
{\emph Physarum polycephalum)~\cite{adamatzkyPhysGate}. The idea behind the gate is that if a travelling localisation (slime mould in~\cite{adamatzkyPhysGate} or excitation wave-fragment in present paper) propagates by itself and does not interact with 
other localisations then the localisation always continues along its original trajectory without changing its velocity vector. 
If two localisations collide their velocity vectors are altered and they exit the gates via different channels. If the first localisation hits 
the refractory tail of a second localisation then the first localisation annihilates. 

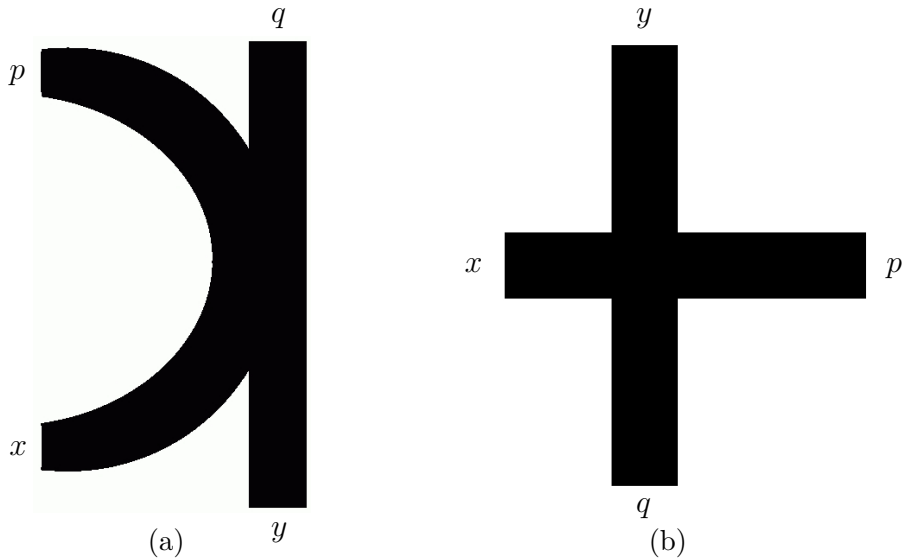
\begin{figure}[!tbp]
	\centering
	\subfigure[]{
		\begin{overpic}[scale=.65]
			{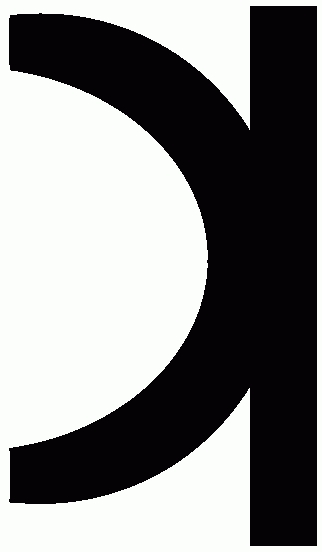}
			\put(-5,12){$x$}
			\put(50,-5){$y$}
			\put(-5,91){$p$}
			\put(50,103){$q$}
		\end{overpic}
	}
	\qquad
	\qquad
	\subfigure[]{
		\begin{overpic}[scale=.3]
			{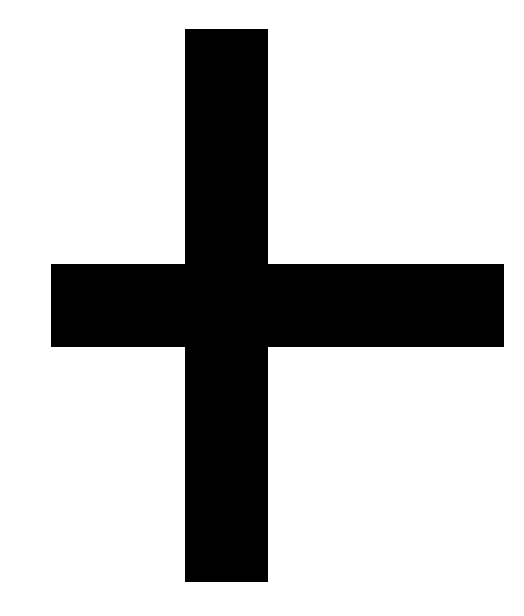}
			\put(0,49){$x$}
			\put(35,100){$y$}
			\put(86,49){$p$}
			\put(35,0){$q$}
		\end{overpic}
	}
	\caption{Geometric structure of gates (a) $P_1$ and (b) $P_2$:  $x$ and $y$ are inputs, $p$ and $q$ are outputs. 
	From~\cite{adamatzkyPhysGate}.}
	\label{gates}
\end{figure}

The geometric structure of gates $P_1$ and $P_2$ are shown in Fig.~\ref{gates}. Input variables are $x$ and $y$ and outputs are $p$ and $q$. Presence of a wave-fragment in a given channel indicates {\sc Truth}, or value '1', 
and absence --- {\sc False}, or value '0'. Each gate implements a transformation from $\left< x, y \right> \rightarrow \left<p, q\right>$. To represent value `1' in input channel $x$ or channel $y$ we generate an excitation near the entrance of the channel. Two wave-fragments are formed. One of the wave fragment travels away from the gate, but the other travels through the channels towards the outputs. The fragment travelling within the channels is only considered to have registered an output when it reaches the end of the respective output channel, we consider the output value in that channel is `1', otherwise if it does not reach the end of the channel we consider the output to be `0'.

To test the full functionality of the gates, we run simulations with every possible input combination $\left<x, y\right> = \left<0, 1\right>$, $\left<1, 0\right>$, and $\left<1, 1\right>$ as well as a range of selected values of $\phi \in A \cup B$, where:
\begin{itemize}
  \item $A = \left\{ 0.07 + n \times 0.001\text{ : }n\text{ is an integer; and }0 \leq n \leq 20 \right\}$
  \item $B = \left\{ 0.077 + n \times 0.0001\text{ : }n\text{ is an integer; and }0 \leq n \leq 20 \right\}$
\end{itemize}
With the value of $\phi$ increasing from the lowest value used, $\phi=0.07$ to the largest value, $\phi=0.09$, 
the excitability of the simulated BZ medium changes from excitable to sub-excitable, and finally non-excitable. 
Therefore with the same input values, different outputs may be generated. The following is a summary of 
the simulation results obtained when using gates $P_1$ and $P_2$.

\subsection{Gate $P_1$}
\begin{figure}[!tbp]
\centering
	\subfigure[]{
		\includegraphics[width=0.22\linewidth]{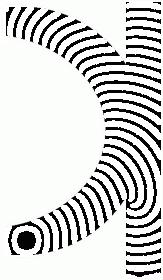}	
	}
	\subfigure[]{
		\includegraphics[width=0.22\linewidth]{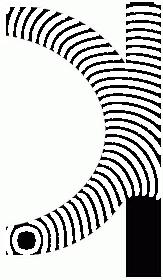}	
	}
	\subfigure[]{
		\includegraphics[width=0.22\linewidth]{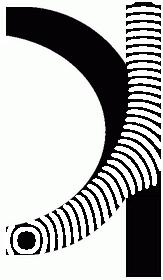}	
	}
	\subfigure[]{
		\includegraphics[width=0.22\linewidth]{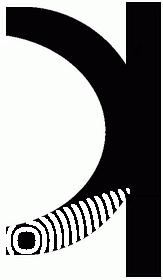}	
	}
\caption{Four possible scenarios observed with gate $P_1$ when input values $\left<x, y\right> = \left<1, 0\right>$. The examples shown here are achieved using specific values of $\phi$: (a) 0.07, (b) 0.076, (c) 0.0777 and (d) 0.079.}
\label{p1x1y0}
\end{figure}

For gate $P_1$, when $\left<x, y \right> = \left<1, 0 \right>$, there are four scenarios observed depending on the value of $\phi$, as shown in Fig.~\ref{p1x1y0}. In the first scenario (Fig.~\ref{p1x1y0}a), when $\phi \leq 0.074$, the wave started in channel $x$ travels to the intersection and enters all three other channels including the other input channel $y$. This is because at this value of $\phi$ the reaction exhibits a high level of excitability. In the second scenario (Fig.~\ref{p1x1y0}b), when $0.075 \leq \phi \leq 0.0771$, the wave reaches the intersection of the junctions but only travels to the two output channels $p$ and $q$. For both of these scenarios the output is $\left<p, q \right> = \left<1, 1 \right>$. However, you could imagine a scenario where this gate with input $\left<x \right>$ only would have outputs $\left<p, q, r \right>$ where output channel $\left<r \right>$ is equivalent to input channel $\left<y \right>$. 

In the third scenario (Fig.~\ref{p1x1y0}c), when $0.0773 \leq \phi \leq 0.0777$, with the excitability of the simulated BZ medium further decreased, only channel $q$ would see an output wave-fragment thus the outputs $\left<p, q\right> = \left<0, 1\right>$. And finally (Fig.~\ref{p1x1y0}d), when $\phi \geq 0.0779$, the wave-fragment cannot travel far from its origin, due to a low level of excitability and also the constraining nature of the channels structure, and the outputs are $\left<p, q\right> = \left<0, 0\right>$.

\begin{figure}[!tbp]
\centering
	\subfigure[]{
		\includegraphics[width=0.22\linewidth]{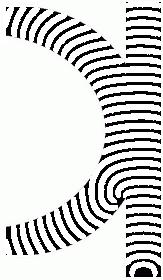}	
	}
	\subfigure[]{
		\includegraphics[width=0.22\linewidth]{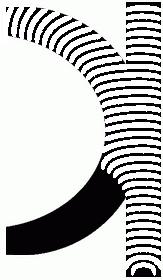}	
	}
	\subfigure[]{
		\includegraphics[width=0.22\linewidth]{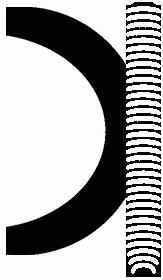}	
	}
	\subfigure[]{
		\includegraphics[width=0.22\linewidth]{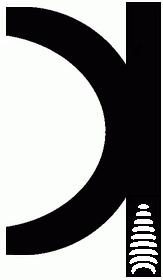}	
	}
\caption{Four scenarios observed using gate $P_1$ when input values $\left<x, y\right> = \left<0, 1\right>$. The examples shown here are achieved using specific values of $\phi$: (a) 0.07, (b) 0.076, (c) 0.0785 and (d) 0.079.}
\label{p1x0y1}
\end{figure}

Similarly, when $\left<x, y\right> = \left<0, 1\right>$, there are also four different scenarios observed in the simulation experiments depending on the different levels of excitability, as show in Fig.~\ref{p1x0y1}. When $\phi \leq 0.074$, the wave-fragment originating from channel $y$ travels into the intersection between the channels and into the three other channels (Fig.~\ref{p1x0y1}a); when $0.075 \leq \phi \leq 0.0772$, the wave fragment travels into the intersection but only travels to the end of the two output channels $p$ and $q$ (Fig.~\ref{p1x0y1}b). Again in both cases the outputs $\left<p, q\right> = \left<1, 1\right>$. 
Whereas in the other two scenarios, when $0.0776 \leq \phi \leq 0.0787$, the outputs $\left<p, q\right> = \left<0, 1\right>$ (Fig.~\ref{p1x0y1}c); and when $\phi \geq 0.0788$, $\left<p, q\right> = \left<0, 0\right>$ (Fig.~\ref{p1x0y1}d). 

\begin{figure}[!tbp]
\centering
	\subfigure[]{
		\includegraphics[width=0.22\linewidth]{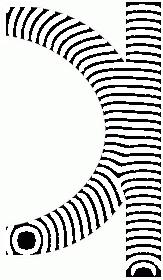}	
	}
	\subfigure[]{
		\includegraphics[width=0.22\linewidth]{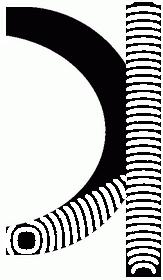}	
	}
	\subfigure[]{
		\includegraphics[width=0.22\linewidth]{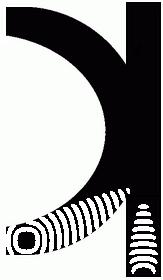}	
	}
	\subfigure[]{
		\includegraphics[width=0.22\linewidth]{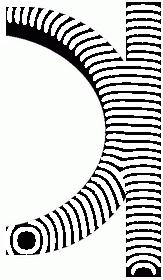}	
	}
\caption{Three scenarios observed when using gate $P_1$ when the input values $\left<x, y\right> = \left<1, 1\right>$ (a-c), and (d) shows an example of an indecisive result. The examples shown here are achieved from specific values of $\phi$: (a) 0.076, (b) 0.0785, (c) 0.079 and (d) 0.0773.}
\label{p1x1y1}
\end{figure}

Finally, when $\left<x, y\right> = \left<1, 1\right>$, only three different scenarios are observed, as shown in Fig.~\ref{p1x1y1}. When $\phi \leq 0.0771$, $\left<p, q\right> = \left<1, 1\right>$ (Fig.~\ref{p1x1y1}a), when $0.0774 \leq \phi \leq 0.0787$ 
$\left<p, q\right> = \left<0, 1\right>$ (Fig.~\ref{p1x1y1}b), and when $\phi \geq 0.0788$, $\left<p, q\right> = \left<0, 0\right>$ (Fig.~\ref{p1x1y1}c).

Not every result of the specified values of $\phi$ are included in the above summary. For example, as shown in Fig.~\ref{p1x1y1}d, when $\left<x, y\right> = \left<1, 1\right>$ and $\phi = 0.0773$, the wave-fragment reached the end of the output channel $p$, but only part of the wave is present due to a lower level of excitability. In this case, it is arguable that the output value in the channel $p$ is `1' or `0', and we consider it as indeterminate and note it with `-'. Thus the output in this particular case is $\left<p, q\right> = \left<-, 0\right>$.

\begin{table}
\centering
\begin{tabular}{c || c | c | c | c | c | c}
\multirow{2}{*}{$\left<x, y\right>$} & 
\multirow{2}{*}{$\phi $} & 
\multirow{2}{*}{$\leq 0.0771$} & 
\multirow{2}{*}{$0.0772$} & 
\multirow{2}{*}{$0.0773$} & 
$0.0774$ & 
$0.0776$ \\ 
 & & & & & $0.0775$ & $0.0777$ \\
\hline
$\left<0, 0\right>$ & 
\multirow{4}{*}{$\left<p, q\right>$} & 
$\left<0, 0\right>$ & 
$\left<0, 0\right>$ & 
$\left<0, 0\right>$ & 
$\left<0, 0\right>$ & 
$\left<0, 0\right>$ \\
\cline{3-7} 
$\left<0, 1\right>$ & 
 & 
$\left<1, 1\right>$ & 
$\left<1, 1\right>$ & 
$\left<-, 1\right>$ & 
$\left<-, 1\right>$ & 
$\left<0, 1\right>$ \\
\cline{3-7} 
$\left<1, 0\right>$ & 
 & 
$\left<1, 1\right>$ & 
$\left<-, 1\right>$ & 
$\left<0, 1\right>$ & 
$\left<0, 1\right>$ & 
$\left<0, 1\right>$ \\
\cline{3-7} 
$\left<1, 1\right>$ & 
 & 
$\left<1, 1\right>$ & 
$\left<-, 1\right>$ & 
$\left<-, 1\right>$ & 
$\left<0, 1\right>$ & 
$\left<0, 1\right>$ \\
\hline
\multicolumn{2}{c|}{channel $p$} & $x \vee y$ & - & - & - & 0\\
\hline
\multicolumn{2}{c|}{channel $q$} & $x \vee y$ & $x \vee y$ & $x \vee y$ & $x \vee y$ & $x \vee y$\\
\end{tabular}

\begin{tabular}{c}
 \\
\end{tabular}
 
\begin{tabular}{c || c | c | c | c }
$\left<x, y\right>$ & 
$\phi $ & 
$0.0778$ & 
$\left[0.0779, 0.0787\right]$ & 
$\geq 0.0788$ \\
\hline
$\left<0, 0\right>$ & 
\multirow{4}{*}{$\left<p, q\right>$} & 
$\left<0, 0\right>$ & 
$\left<0, 0\right>$ & 
$\left<0, 0\right>$ \\
\cline{3-5} 
$\left<0, 1\right>$ & 
 & 
$\left<0, 1\right>$ & 
$\left<0, 1\right>$ & 
$\left<0, 0\right>$ \\
\cline{3-5} 
$\left<1, 0\right>$ & 
 & 
$\left<0, -\right>$ & 
$\left<0, 0\right>$ & 
$\left<0, 0\right>$ \\
\cline{3-5} 
$\left<1, 1\right>$ & 
 & 
$\left<0, 1\right>$ & 
$\left<0, 1\right>$ & 
$\left<0, 0\right>$ \\
\hline
\multicolumn{2}{c|}{channel $p$} & 0 & 0 & 0\\
\hline
\multicolumn{2}{c|}{channel $q$} & - & $y$ & $0$\\
\end{tabular}

\caption{Functionality of gate $P_1$ at different value ranges of $\phi$ ($y$-dominant).}
\label{tablep1}
\end{table}

By combining the above results for all possible input values together, we are able to determine the functionality of the gate at different values of $\phi$, as shown in Tab.~\ref{tablep1}. In channel $p$, one can implement the $x \vee y$ logic, and in channel $q$, one can implement both the $x \vee y$ logic and the $y$ logic.

\begin{figure}[!tbp]
\centering
	\subfigure[]{
		\includegraphics[width=0.22\linewidth]{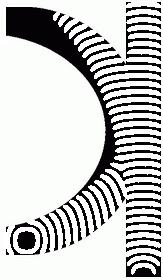}	
	}
	\qquad
	\qquad
	\qquad
	\subfigure[]{
		\includegraphics[width=0.22\linewidth]{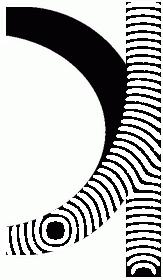}	
	}
\caption{Gate $P_1$ has different functionality depending on whether: (a)~$y$-dominant and (b)~$x$-dominant. This is determined by which wave-fragment (originating from channel $x$ or $y$) arrives at the junction first.}
\label{p1domination}
\end{figure}

Furthermore, on closer inspection to Tab.~\ref{tablep1}, one may find that for the input values of $\left<x, y\right> = \left<0, 1\right>$ and $\left<x, y\right> = \left<1, 1\right>$, the output values are almost identical in every range of $\phi$. This is what we call a $y$-dominant situation. In addition, there is an $x$-dominant situation. The reason why it is $x$-dominant or $y$-dominant, may be seen by studying the time lapse images shown in Fig.~\ref{p1domination}. Both images are achieved at $\phi = 0.0779$, the difference is that in Fig.~\ref{p1domination}b the origin of the wave in input channel $x$ is farther away from the edge of the channel than in Fig.~\ref{p1domination}a. And when the two wave-fragments collide and join together, it looks as if the wave-fragment from channel $x$ is absorbed, thus $y$-dominant, in Fig.~\ref{p1domination}a. And in Fig.~\ref{p1domination}b, it looks as if the wave-fragment from channel $y$ is absorbed, thus $x$-dominant. In this case, whether the gate $P_1$ appears to be $x$-dominant or $y$-dominant is determined by which wave-fragment (originating from channel $x$ or $y$) arrives at the junction first.

It is evident that in an $x$-dominant situation, the functionality of gate $P_1$ (summarised in Tab.~\ref{tablep1-2}) is different from that when there is a $y$-dominant situation. It shows that for the $x$-dominant situation, one can implement a $\neg x \wedge y$ logic.

\begin{table}
\centering
\begin{tabular}{c || c | c | c | c | c }
\multirow{3}{*}{$\left<x, y\right>$} & 
\multirow{3}{*}{$\phi $} & 
\multirow{3}{*}{$\leq 0.0771$} & 
\multirow{3}{*}{$0.0772$} & 
$0.0773$ & 
\multirow{3}{*}{$\left[0.0776, 0.0779 \right]$} \\
 & & & & $0.0774$ & \\
 & & & & $0.0775$ & \\
\hline
$\left<0, 0\right>$ & 
\multirow{4}{*}{$\left<p, q\right>$} & 
$\left<0, 0\right>$ & 
$\left<0, 0\right>$ & 
$\left<0, 0\right>$ & 
$\left<0, 0\right>$ \\
\cline{3-6} 
$\left<0, 1\right>$ & 
 & 
$\left<1, 1\right>$ & 
$\left<1, 1\right>$ & 
$\left<-, 1\right>$ & 
$\left<0, 1\right>$ \\
\cline{3-6} 
$\left<1, 0\right>$ & 
 & 
$\left<1, 1\right>$ & 
$\left<-, 1\right>$ & 
$\left<0, 1\right>$ & 
$\left<0, 1\right>$ \\
\cline{3-6} 
$\left<1, 1\right>$ & 
 & 
$\left<1, 1\right>$ & 
$\left<-, 1\right>$ & 
$\left<0, 1\right>$ & 
$\left<0, 1\right>$ \\
\hline
\multicolumn{2}{c|}{channel $p$} & $x \vee y$ & - & - & 0\\
\hline
\multicolumn{2}{c|}{channel $q$} & $x \vee y$ & $x \vee y$ & $x \vee y$ & $x \vee y$\\
\end{tabular}

\begin{tabular}{c}
 \\
\end{tabular}
 
\begin{tabular}{c || c | c | c | c }
$\left<x, y\right>$ & 
$\phi $ & 
$0.078$ & 
$\left[0.0781, 0.0787\right]$ & 
$\geq 0.0788$ \\
\hline
$\left<0, 0\right>$ & 
\multirow{4}{*}{$\left<p, q\right>$} & 
$\left<0, 0\right>$ & 
$\left<0, 0\right>$ & 
$\left<0, 0\right>$ \\
\cline{3-5} 
$\left<0, 1\right>$ & 
 & 
$\left<0, 1\right>$ & 
$\left<0, 1\right>$ & 
$\left<0, 0\right>$ \\
\cline{3-5} 
$\left<1, 0\right>$ & 
 & 
$\left<0, -\right>$ & 
$\left<0, 0\right>$ & 
$\left<0, 0\right>$ \\
\cline{3-5} 
$\left<1, 1\right>$ & 
 & 
$\left<0, -\right>$ & 
$\left<0, 0\right>$ & 
$\left<0, 0\right>$ \\
\hline
\multicolumn{2}{c|}{channel $p$} & 0 & 0 & 0\\
\hline
\multicolumn{2}{c|}{channel $q$} & - & $\neg x \wedge y$ & $0$\\
\end{tabular}

\caption{Functionality of gate $P_1$ at different value ranges of $\phi$ ($x$-dominant).}
\label{tablep1-2}
\end{table}

\subsection{Gate $P_2$}

For gate $P_2$, there were three different scenarios observed. The first one is when $\phi \leq 0.075$ and the simulated BZ medium is excitable, at the junction wave-fragments travel to all possible routes and arrive in both output channels $p$ and $q$ in addition to free input channel, as shown in Fig.~\ref{p2-1}. Note that in Fig.~\ref{p2-1}c, the wave-fragment originating from channel $x$ arrives at the junction first and travels into channel $y$, where it collides with the wave-fragment originating from channel $y$ and the wave fragment originating from $y$ is annihilated.

\begin{figure}[!tbp]
\centering
	\subfigure[]{
		\includegraphics[width=0.28\linewidth]{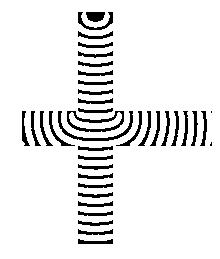}	
	}
	\subfigure[]{
		\includegraphics[width=0.28\linewidth]{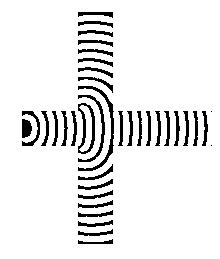}	
	}
	\subfigure[]{
		\includegraphics[width=0.28\linewidth]{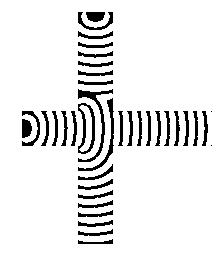}	
	}
\caption{Functionality of Gate $P_2$ when the simulated Bz medium is excitable. Wave-fragments would arrive at both output channels $p$ and $q$, thus $\left<p, q\right> = \left<1, 1\right>$: (a) $\left<x, y\right> = \left<0, 1\right>$, (b) $\left<x, y\right> = \left<1, 0\right>$ and (c) $\left<x, y\right> = \left<1, 1\right>$. In all three examples, $\phi = 0.070$.}
\label{p2-1}
\end{figure}

In the second scenario, when $0.076 \leq \phi \leq 0.0787$ and the simulated BZ medium is sub-excitable, the wave-fragment travels straight through the junction to the opposite output channel when there is only one wave-fragment, as shown in Fig.~\ref{p2-2}a and Fig.~\ref{p2-2}b. When $\left<x, y\right> = \left<1, 1\right>$, when the wave-fragment originating from channel $y$ arrives at the junction, it would collide with the tail of the wave-fragment originating from channel $x$ and subsequently be annihilated, as shown in Fig.~\ref{p2-2}c. This is due to the relative lengths of the input channels prior to reaching the junction.

\begin{figure}[!tbp]
\centering
	\subfigure[]{
		\includegraphics[width=0.28\linewidth]{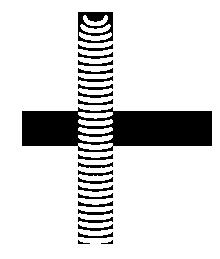}	
	}
	\subfigure[]{
		\includegraphics[width=0.28\linewidth]{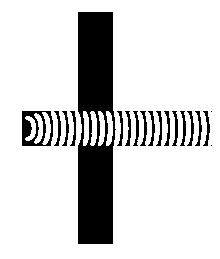}	
	}
	\subfigure[]{
		\includegraphics[width=0.28\linewidth]{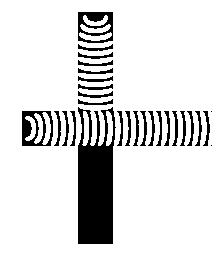}	
	}
\caption{Functionality of gate $P_2$ when the simulated medium is close to a sub-excitable state. Wave-fragments are observed to travel straight through the junction to the opposite output channel where possible. The outputs $\left<p, q\right>$ are either $\left<1, 0\right>$ or $\left<0, 1\right>$: (a) $\left<x, y\right> = \left<0, 1\right>$, (b) $\left<x, y\right> = \left<1, 0\right>$ and (c) $\left<x, y\right> = \left<1, 1\right>$. In all three examples, $\phi = 0.0785$.}
\label{p2-2}
\end{figure}

In the third scenario, when $\phi \geq 0.0788$ and the simulated BZ medium is unexcitable, no wave-fragments are observed to arrive at the output channels $p$ and $q$ since they collapse not long after being initiated. Therefore the outputs $\left<p, q\right> = \left<0, 0\right>$.

\begin{figure}[!tbp]
\centering
	\subfigure[]{
		\includegraphics[width=0.28\linewidth]{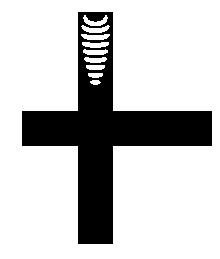}	
	}
	\subfigure[]{
		\includegraphics[width=0.28\linewidth]{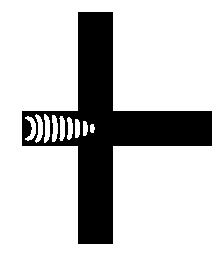}	
	}
	\subfigure[]{
		\includegraphics[width=0.28\linewidth]{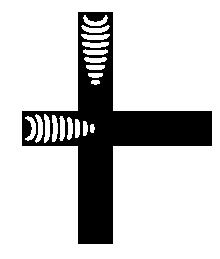}	
	}
\caption{Functionality of gate $P_2$ when the simulated BZ medium is in an unexcitable state. Wave-fragments collapse prior to interacting at the junction therefore no excitation waves reach the output channels $p$ and $q$, thus $\left<p, q\right> = \left<0, 0\right>$: (a) $\left<x, y\right> = \left<0, 1\right>$, (b) $\left<x, y\right> = \left<1, 0\right>$ and (c) $\left<x, y\right> = \left<1, 1\right>$. In all three examples, $\phi = 0.079$.}
\label{p2-3}
\end{figure}

The functionality of Gate $P_2$ with different levels of excitability is summarised in Table~\ref{tablep2}.

\begin{table}
\centering
\begin{tabular}{c || c | c | c | c }
$\left<x, y\right>$ & 
$\phi $ & 
$\leq 0.075$ & 
$\left[0.076, 0.0787\right]$ & 
$\geq 0.0788$ \\
\hline
$\left<0, 0\right>$ & 
\multirow{4}{*}{$\left<p, q\right>$} & 
$\left<0, 0\right>$ & 
$\left<0, 0\right>$ & 
$\left<0, 0\right>$ \\
\cline{3-5} 
$\left<0, 1\right>$ & 
 & 
$\left<1, 1\right>$ & 
$\left<0, 1\right>$ & 
$\left<0, 0\right>$ \\
\cline{3-5} 
$\left<1, 0\right>$ & 
 & 
$\left<1, 1\right>$ & 
$\left<1, 0\right>$ & 
$\left<0, 0\right>$ \\
\cline{3-5} 
$\left<1, 1\right>$ & 
 & 
$\left<1, 1\right>$ & 
$\left<1, 0\right>$ & 
$\left<0, 0\right>$ \\
\hline
\multicolumn{2}{c|}{channel $p$} & $x \vee y$ & x & 0\\
\hline
\multicolumn{2}{c|}{channel $q$} & $x \vee y$ & $\neg x \wedge y$ & $0$\\
\end{tabular}
\caption{}
\label{tablep2}
\end{table}

\section{Simulated one-bit half-adder}
\label{simulatedadder}

\begin{figure}[!tbp]
\centering
\includegraphics[width=0.6\textwidth]{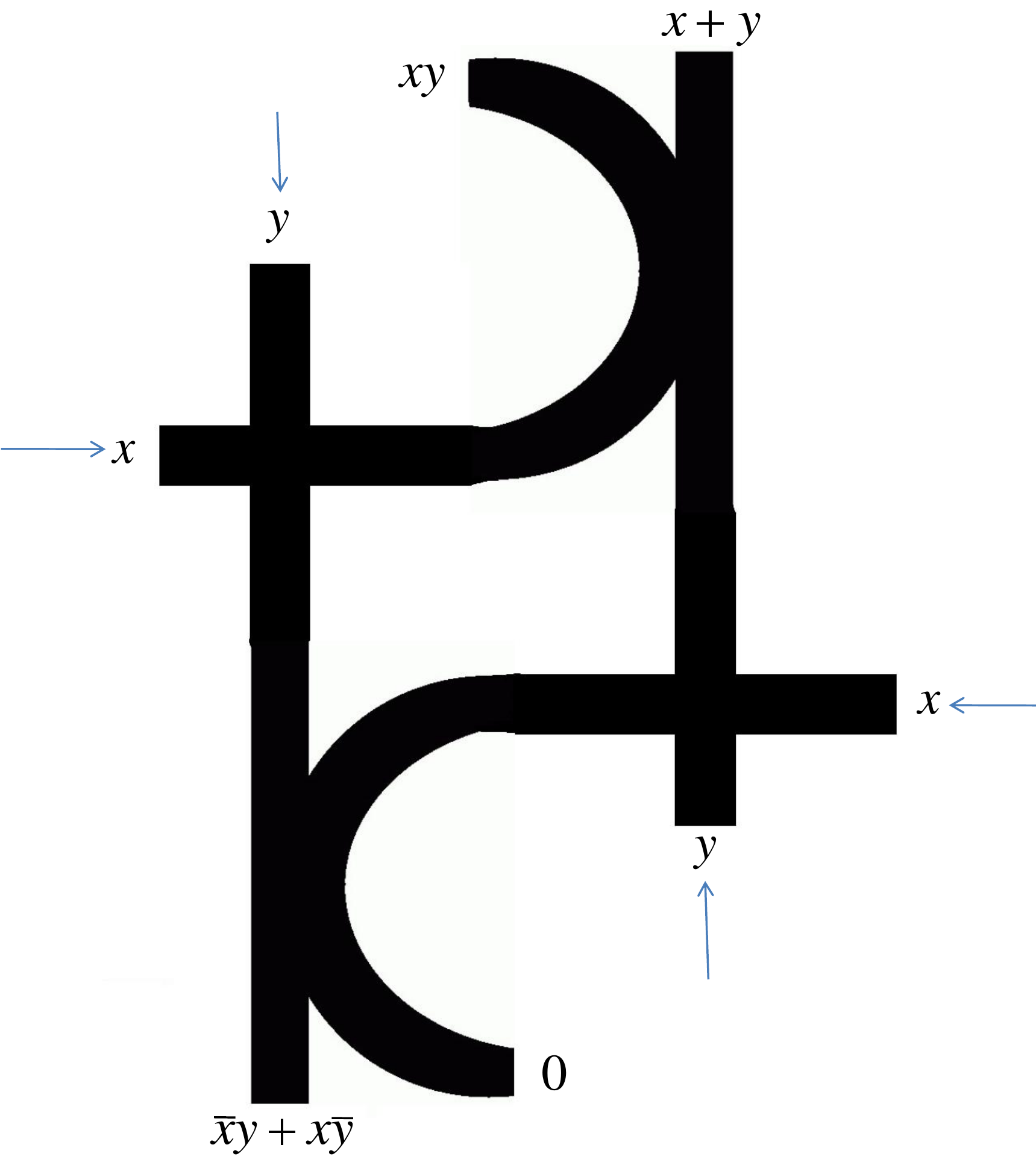}
\caption{Scheme of one-bit half-adder constructed from gates $P_1$ and $P_2$. Inputs are indicated by arrows. Outputs 
$\overline{x}y + x\overline{y}$ and $xy$ are sum and carry values computed by the adder. Outputs $0$ and 
$x+y$ are byproducts.}
\label{adderscheme}
\end{figure}

A one-bit half-adder is a logical circuit which takes two inputs $x$ and $y$ and produces two outputs: 
sum $\overline{x}y + x\overline{y}$ and carry $xy$. In previous work a one-bit half-adder was constructed with Physarum and required two copies of gate $P_1$ (Fig.~\ref{gates}a) and two copies of gate $P_2$ (Fig.~\ref{gates}b).
Cascading the gates into the adder is shown in Fig.~\ref{adderscheme}. Signals $x$ and $y$ are inputted using the two $P_2$ gates. 
The outputs of the two $P_2$ gates are connected to the inputs of the two $P_1$ gates.

\begin{figure}[!tbp]
\centering
	\subfigure[]{
		\includegraphics[width=0.45\linewidth]{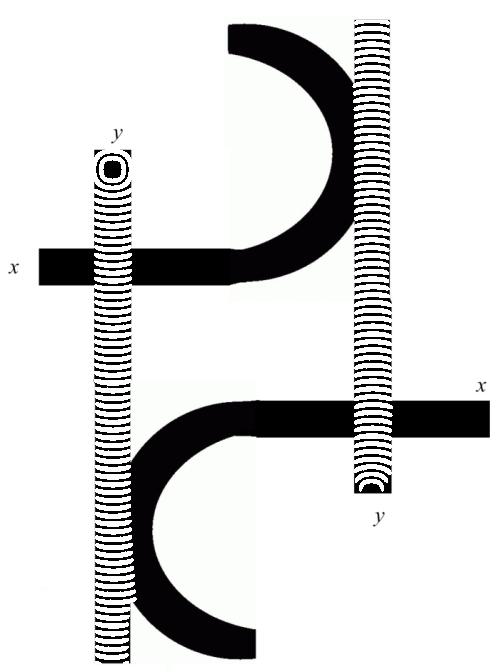}	
	}
	\subfigure[]{
		\includegraphics[width=0.45\linewidth]{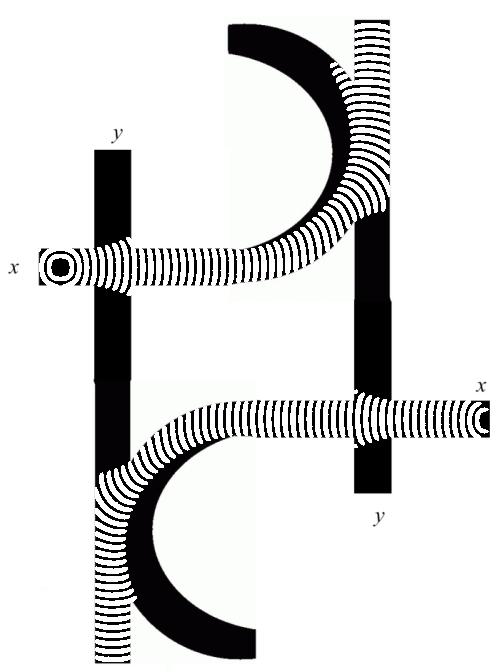}	
	}
	\subfigure[]{
		\includegraphics[width=0.45\linewidth]{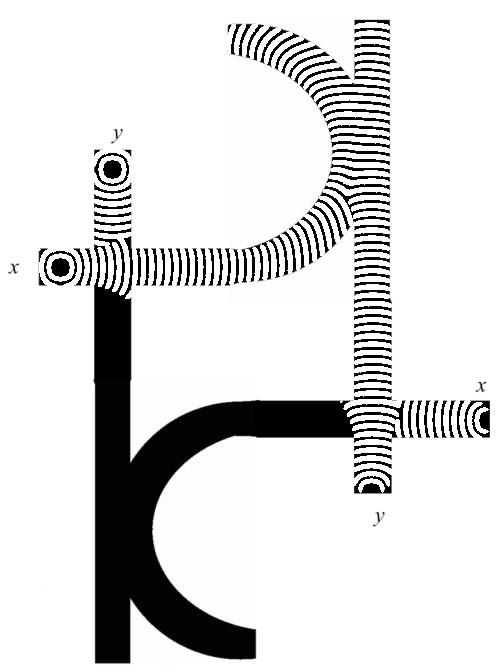}	
	}
\caption{Time lapsed images of localized wave-fragments traveling in the one-bit half-adder scheme in the simulated BZ medium. Dynamics of excitation
is shown for the following input values and parameters: (a)~$\left<x, y\right> = \left<0, 1\right>, \phi=0.0785$, 
(b)~$\left<x, y\right> = \left<1, 0\right>, \phi=0.0774$, (c)~$\left<x, y\right> = \left<1, 1\right>, \phi=0.0770$}
\label{adder}
\end{figure}

Simulation of the adder using the Oregonator model modified for light sensitivity was successful (Fig.~\ref{adder}). To simulate inputs $x=0$ and $y=1$ we initiate wave-fragments at the beginning of the $y$ input channels (marked $y$ and arrow in Fig.~\ref{adderscheme}). The wave-fragments propagate along these channels. The waves do not branch at the junctions with other channels because we keep the wave-fragments localized by varying the parameter $\phi$ (Fig.~\ref{adder}ab).

For input values $x=1$ and $y=0$ wave-fragments are initiated at sites marked $x$ and arrow in Fig.~\ref{adderscheme}. The wave-fragment initiated in the left $x$-input channel propagates towards the $x+y$-output of the adder. The wave-fragment
initiated in the right $x$-input channel travels towards $\overline{x}y + x \overline{y}$ (Fig.~\ref{adder}).

When both inputs are activated, $x=1$ and $y=1$, the wave-fragment originating from the left $y$-input channel is 
blocked by the refractory tail of the wave-fragment originating from the left $x$-input channels. The wave-fragment traveling in the right $x$-input channel is blocked by the refractory tail of the wave-fragment traveling in the right $y$-input channel. The wave-fragments representing $x=1$ and $y=1$ enter gate $P_1$ at the top-right-hand-side of the scheme and emerge at its outputs $xy$ and $x+y$ (Fig.~\ref{adder}). Thus the functionality of the designed circuit Fig.~\ref{adder} is demonstrated in computer simulation. It is interesting to note that there is a specific value of $\phi$ required in order to observe the correct function of the adder for each possible input sequence. This highlights the high degree of controllability that can be implemented when using computer simulations. It also highlights in line with the results obtained for the single gates how sensitively dependent the functionality of the gate is on the level of excitability conferred by $\phi$. If a single value of $\phi$ was used then certain functionality could not be achieved.

\section{Experimental techniques}
\label{experimentaltechniques}
 
Sodium bromate, sodium bromide, malonic acid, sulphuric acid, tris(bipyridyl) ruthenium(II) chloride, 27\% sodium silicate solution stabilized in 4.9 M sodium hydroxide were purchased from (Sigma-Aldrich, U.K., BH12 4QH) and used as received unless stated otherwise.

To create the gels a stock solution of the sodium silicate was prepared by mixing 222ml of the purchased sodium silicate solution with 57 ml of 2 M sulphuric acid and 180 ml of deionised water. Ru(bpy)$_3$SO$_4$ was recrystalised from the choloride salt with sulphuric acid. Solutions for making the gels were prepared by mixing 2.5 ml of the acidified silicate solution and 0.6ml of 0.025M Ru(bpy)$_3$SO$_4$ with 0.7 ml of 1M sulphuric acid solution. Using capillary action, portions of this solution were quickly transferred into a custom-designed 25cm long 0.3 mm deep Perspex mould covered with microscope slides. The solution was left for 3 hours to permit complete gellation. After gellation the adherence to the Perspex mould is negligible leaving a thin layer on the glass slide. After 3 hours the slides were carefully removed from the mould and the gels on the slides were washed in deionised water at least five times. The gels were 26 mm by 26mm, with a wet thickness of approximately 300$\mu$m. The gels were stored under water and rinsed with deionised water just before use.

The catalyst free reaction mixture was freshly prepared in a 30 ml continuously-fed stirred tank reactor (CSTR), which involved the in situ synthesis of stoichiometric bromomalonic acid from malonic acid and bromine generated from the partial reduction of sodium bromate. This CSTR in turn continuously fed a thermostatted open reactor with fresh catalyst-free BZ solution in order to maintain a nonequlibrium state. The final composition of the catalyst-free reaction solution in the reactor was: 0.42 M sodium bromate, 0.19 M malonic acid, 0.52 M sulphuric acid and 0.11 M bromide. The residence time was 30 minutes. 

The open reactor was surrounded by a water jacket thermostatted at 20$^\circ$C. Peristaltic pumps (Watson Marlow Ltd. UK. TR11 4RU) were used to pump the reaction solution into the reactor and remove the effluent. A diagrammatic representation of the experimental setup is shown in Fig.~\ref{Fig1}.

A Sanyo PROxtrax multiverse projector (Sanyo UK WD24 4PT) was used to illuminate the computer-controlled image. Images were captured using a Lumenera infinity 2 USB 2.0 scientific digital camera fitted with a macro video zoom lens (18-108 f/2.5) and a blue filter (Edmunds optics, UK. YO26 6BL). The computer generated images of the gates were projected onto the catalyst loaded gel via an optically flat mirror (Edmunds optics). The spatially distributed excitable field on the surface of the gel was achieved by the projection of dark coloured channels on a light coloured background. The light intensity of the channels was controlled via in house written software. The size of the projected channel width was approximately 1.7 mm. Every 10 seconds, the pattern was replaced with a uniform light level of 5.7mW cm$^-$$^2$ for 10 ms during which time an image of the BZ fragments on the gel was captured. The purpose of removing the projected image during this period was to allow activity on the gel to be more visible to the camera.

Captured images were processed to identify chemical activity. This was done by differencing successive images to create a black and white image. The levels of excitation in the analysed cells are shown in white and the background in black. The images were cropped and the channel boundary was superimposed on the images to aid analysis of the results.

\begin{figure}[!tbp] 
\centering
\includegraphics[width=0.75\linewidth]{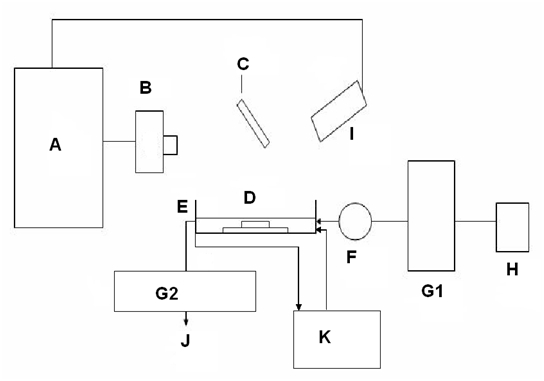}	
\caption{A block diagram of the experimental setup where A: computer, B: projector, C: mirror, D: microscope slide with the catalyst-laden gel, E: thermostatted Petri dish, F: CSTR, G1 and G2: pumps, H: stock solutions, I: camera, J: effluent flow, K: thermostatted water bath.}
\label{Fig1}
\end{figure}

\section{Experimental results}
\label{experimentalresults}

The scheme for the half bit adder had been successfully designed and implemented using the Oregonator model of the light sensitive BZ reaction. The half adder scheme consists of two pairs of interaction gates (Fig.~\ref{gates}a,b) which must be solved in order to successfully implement the half-adder. Therefore, because of the complex nature of the interactions and their high dependency on synchronised inputs (more challenging in experimental implementation) a strategy of splitting the one-bit half-adder into its constituent parts and solving each part separately was devised. This involved projecting only parts of the adder scheme onto the thin film light sensitive BZ reaction and injecting fragments into the relevant inputs. By varying the light levels in the channels the full range of behaviour of the interaction gates could be explored. This enabled the optimal light level for solving each gate to be identified. Then using the optimal conditions an attempt to construct the one-bit half-adder scheme in experiment was undertaken.    

\subsection{Results of implementing interaction Gate $P_1$}

\begin{figure}[!tbp]
\centering
\includegraphics[width=0.45\linewidth]{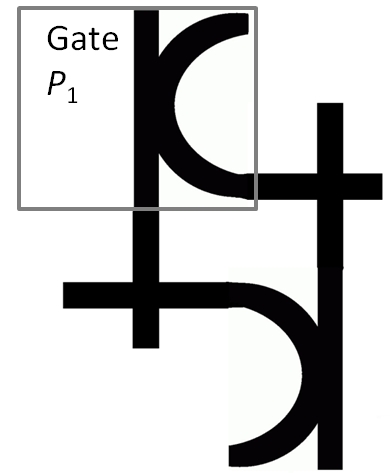}
\caption{Structure of the one-bit half-adder showing the position of the interaction gate $P_1$ within the overall scheme.}
\label{adderP1}
\end{figure}

The geometric structure of gate $P_1$ where $x$ and $y$ are inputs channels and $p$ and $q$ are output channels is shown in
Fig.~\ref{gates}a. Also shown in Figure~\ref{adderP1} is the adder scheme identifying the part of the overall adder which constitutes $P_1$. The structures were drawn in mirror image compared to the gate structures shown for the simulation results because in experiment the projected image was reflected on to the gel using a mirror.

\begin{figure}[!tbp] 
\centering
	\subfigure[]{
		\includegraphics[height=26mm]{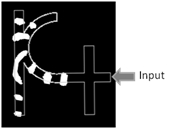}	
	}
	\subfigure[]{
		\includegraphics[height=26mm]{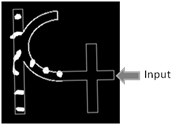}	
	}
	\subfigure[]{
		\includegraphics[height=26mm]{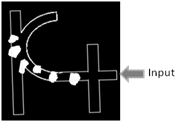}	
	}
\caption{Progression of fragments seen every 30 seconds when $x=1$ at various light intensities where (a)~excitable at 2.1~mW cm$^-$$^2$ (b)~weakly excitable at 2.6~mW cm$^-$$^2$ (c)~sub excitable at 3.1mW cm$^-$$^2$.}
\label{fig12}
\end{figure}

In the chemical experiments using the light sensitive BZ reaction and projecting the image of gate $P_1$, 
when $\langle x,y \rangle = \langle 1,0 \rangle$ there are three different types of behaviour observed for the interaction of the input fragment at the junctions that are sensitively dependent on the excitability level of the channels. The excitability level of the channels is directly correlated to the projected light intensity. For correct implementation of the half adder scheme where $x=1$ there should only be a fragment travelling into output channel $q$. In Fig.~\ref{fig12}a, the light intensity was 
2.1mW cm$^{-2}$ meaning that the channels were excitable and therefore, the input fragment from channel $x$ is able to enter both output channels ($p$ and $q$) but in addition can also travel into the other input channel ($y$).  In Fig.~\ref{fig12}b the light level was increased to 2.6mW cm$^{-2}$ making the channels weakly excitable. The input fragment travelled into the output $q$ channel and input $y$ channel only. This is in contrast to the results observed during the simulation experiments where the fragment at an intermediate light level ($\phi$ value 0.076 in the model) travelled only to the output channels $p$ and $q$. This difference in observed behaviour is due to a decrease in stability of the fragment in the experiment with increasing distance from the initiation site. Therefore, at the first junction the fragment expands and is able to transfer into the free input channel. However, when the fragment reaches the second junction it has lost stability and become localised. The fragment can be seen to follow the contour of the straight output channel.

In Fig.~\ref{fig12}c, the light intensity was increased to 3.1mW cm$^{-2}$ meaning the channels were close to the sub-excitable limit. At this light level the fragment from the input $x$ channel only travelled into the output channel $q$ at the junction between $p$ and $q$. At this light level the fragments are very weak, diffuse and relatively difficult to observe with the camera. Therefore, the differenced image does not show the fragment travelling through the entire output channel.

\begin{figure}[!tbp] 
\centering
		\includegraphics[height=35mm]{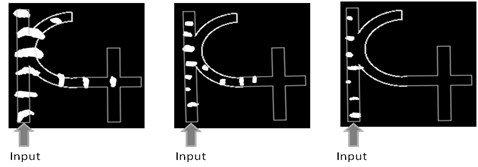}
\caption{Progression of fragments seen every 30 seconds when $y=1$ at various light intensities where (a) excitable at 0.4mW cm$^-$$^2$ (b) weakly excitable at 2.6mW cm$^-$$^2$ (c) sub excitable at 3.1mW cm$^-$$^2$.}
\label{fig13}
\end{figure}

When $\langle x,y \rangle$ = $\langle 0,1 \rangle$ there were also three distinct types of behaviour observed at different light levels. The desired behaviour in order to implement the adder successfully was that an input fragment in the $y$ channel only would result in a single fragment in the $q$ output channel.  In Fig.~\ref{fig13}a the light intensity was 0.4mW cm$^{-2}$ in the excitable domain and again this resulted in the input fragment travelling in the $y$ channel splitting into both output channels $p$ and $q$ and also into the input channel $x$.  In Fig.~\ref{fig13}b the light level was increased to 2.6mW cm$^{-2}$ in a weakly excitable domain and the fragment travelling in the $y$ input channel split at the junction and travelled into the $q$ output channel and the $x$ input channel.  Again this is in contrast to the simulation results where the fragments travelled to the output $p$ and $q$ channels only at the intermediate light level. Finally in Fig.~\ref{fig13}c the light intensity was 
3.1mW cm$^{-2}$ the channels were sub excitable and the fragment travelled only into the output $q$ channel.

\begin{figure}[!tbp] 
\centering
		\includegraphics[height=35mm]{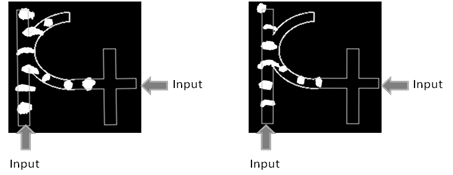}
\caption{Experimental results when $X=1$ and $Y=1$ at 2 different light intensities 
(a) 2.1mW cm$^-$$^2$ and(b) 3.1mW cm$^-$$^2$.}
\label{fig14}
\end{figure}

When $\langle x,y \rangle = \langle 1,1 \rangle$ the desired behaviour (in order to solve the half adder) is for the fragments to collide and merge at the intersection of the output channels and then split to give a fragment in both the $p$ and $q$ output channels. At excitable light levels up to a value of 2.1mW cm$^{-2}$ the fragments travelled from the input $x$ and $y$ channels collided, merged and split at the junction to give fragments at the output $p$ and $q$ channels (Fig.~\ref{fig14}a). However, if the light level was increased to 3.1mW cm$^{-2}$ close to the sub-excitable domain then despite collision of the input fragments a fragment was only observed to travel into the $q$ output channel (Fig.~\ref{fig14}b).  Therefore, despite the optimal light level for solving the $\langle x,y \rangle = \langle 1,0 \rangle$ and $\langle x,y \rangle = \langle 0,1 \rangle$ cases being 
3.1mWcm$^{-2}$ this is different to the light level found to be optimal for solving the $\langle x,y \rangle = \langle 1,1 \rangle$ case. This may have implications for solving the adder using the current scheme if the light level is held at a constant value.

Despite the fact that this interaction gate is part of a larger adder scheme the results obtained in experiment for the 
$\langle x,y \rangle = \langle 1,0 \rangle$ or $\langle 0,1 \rangle$ highlight the adaptability of using light to impart different functionality to logic gates. For both the $x=1$ and $y=1$ case the same general behaviour is observed. Low light intensity (excitable), $x=1$ or $y=1$ output 1, 1 and 1. Medium light intensity (weakly excitable), $x=1$ or $y=1$, output 1,1 and 0. 
High light intensity (sub-excitable), $x=1$ or $y=1$, output 1,0 and 0. Although the fourth case is not investigated it is obvious that if the light level is increased further then the $x=1$ or $y=1$, output 0,0 and 0 could be obtained in experiment. Also in our experimental results we count any excitation at the end of the output channel as 1. However, as demonstrated for the simulation experiments there can also be an indeterminate result. This could be achieved in experiment by setting a threshold range for the level of excitation detected at the output channel. In this way the functionality of the gate could be further increased at differing light levels. In experiment it is not as simply to map all possible light levels because the system of changing the light level does not have as many degrees of freedom as changing parameter $\phi$ in the simulation experiments.      

\subsection{Experimental results for interaction Gate $P_2$}

\begin{figure}[!tbp]
\centering
\includegraphics[width=0.45\linewidth]{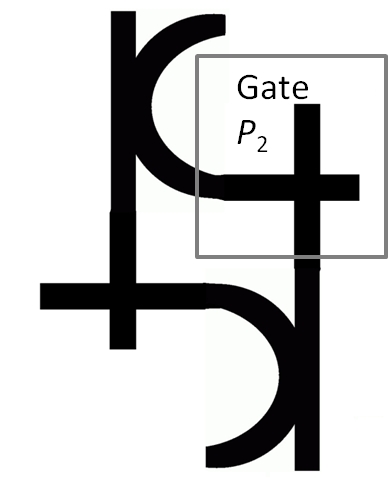}
\caption{Adder structure showing the position of interaction gate $P_2$.}
\label{adderP2}
\end{figure}

The geometric structure of gate $P_2$ where $x$ and $y$ are input channels and $p$ and $q$ are output channels is shown 
in Fig.~\ref{gates}b. The adder structure showing the position of gate $P_2$ is shown in Fig.~\ref{adderP2}. During these experiments only a single section of the adder scheme is solved where the $y$ input channel is longer than $x$. In the other half of the adder scheme this is reversed and the $x$ input channel is longer than $y$. For gate $P_2$ to be solved correctly an optimal light level has to be found so that a fragment in either the $x$ or $y$ channel will travel straight into either the $q$ or $p$ output channels respectively. When there are inputs at both $x$ and $y$ the dominant fragment (from the short input channel, in this case x) should block the fragment from the longer input channel (in this case $y$). The resulting collision at the junction should result in a fragment travelling into the $p$ output channel (in this case).

\begin{figure}[!tbp] 
\centering
		\includegraphics[height=35mm]{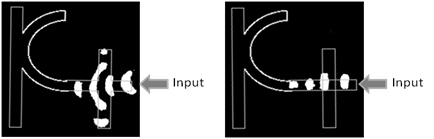}
\caption{Progression of fragment every 30 seconds when $x=1$ at the cross junction (a) excitable at 2.2mW cm$^-$$^2$  (b) sub-excitable 3.1mW cm$^-$$^2$.}
\label{fig16}
\end{figure}

For gate $P_2$ it was found in experiment that at a lower light intensity of 2.2mW cm$^{-2}$ for the
$\langle x,y \rangle = \langle 1,0 \rangle$ case  the fragment from the input $x$ travels to the input $y$ 
and both output channels $p$ and $q$ see Fig.~\ref{fig16}a. This is because at this light level the 
channels are in an excitable domain. However, if the light intensity is raised to 3.1mW cm$^{-2}$  
close to the sub-excitable limit then the fragment from the input $x$ channel only travels to the output $p$ 
channel (Fig.~\ref{fig16}b).

\begin{figure}[!tbp] 
\centering
\subfigure{}{\includegraphics[width=0.4\textwidth]{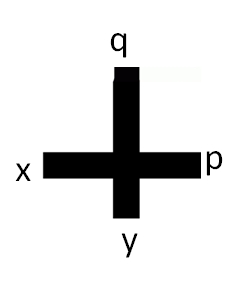}}
\subfigure{}{\includegraphics[width=0.4\textwidth]{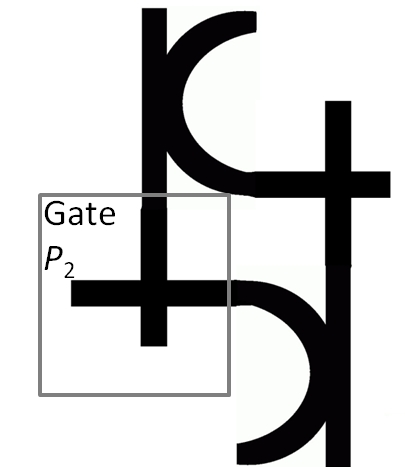}}
\caption{Geometric structure of gate (a)~$P_2'$, where $x$ and $y$ are inputs, $p$ and $q$ are outputs, 
(b)~Adder structure showing the position of gate $P_2'$.}
\label{fig17}
\end{figure}

\begin{figure}[!tbp] 
\centering
		\includegraphics[height=35mm]{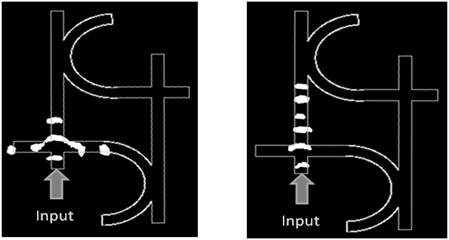}
\caption{Progression of fragments every 30 seconds when $y=1$ at the cross junction at (a)~3.7mW cm$^-$$^2$  and (b)~4.4mW cm$^-$$^2$.}
\label{fig18}
\end{figure}

For the case $\langle x,y \rangle = \langle 0,1 \rangle$ the interaction gate $P_2'$ was used (Fig.~\ref{fig17}a),
note also the position in the overall adder scheme in Fig.~\ref{fig17}b. At a lower light intensity of 3.7mW cm$^{-2}$ 
close to the sub-excitable limit the fragments from the input $y$ channel travels into the input $x$ channel and both output channels $p$ and $q$ (Fig.~\ref{fig18}a). If the light level is raised further to 4.4 mW cm$^{-2}$ then the fragment only travels into the $q$ output channel (Fig.~\ref{fig18}b). The light levels are different from the previous experimental results 
for $\langle x,y \rangle = \langle 1,0 \rangle$ (Fig.~\ref{fig16}ab) due to subtle differences in gel composition and morphology which can make the sub-excitable region variable in experimental implementations. However, the sub-excitable region can always be identified by the formation of unbounded fragments in a projected uniform light level prior to the projection of the interaction gates/channels.

\begin{figure}[!tbp] 
\centering
		\includegraphics[height=35mm]{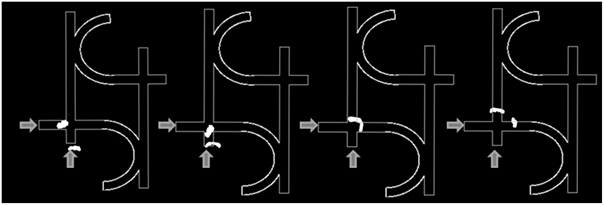}
\caption{Progression of fragment every 10 seconds when $x=1$ and $y=1$ at light intensity 1.5mW cm$^-$$^2$  (excitable) (a) $x$ fragment is dominant (b) this results in a collision and the annihilation of the $y$ fragment  (c) the resulting daughter fragment spreads across the north and east channels (d)~finally it  breaks into 2 fragments which travel along the north and east channels (grey arrows show direction of fragment input).}
\label{fig19}
\end{figure}

\begin{figure}[!tbp] 
\centering
		\includegraphics[height=35mm]{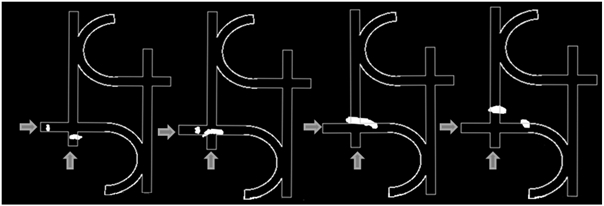}
\caption{Progression of fragment every 10 seconds when $x=1$ and $y=1$ at light intensity 3.0mW cm$^-$$^2$  (weakly excitable)
(a)~$y$ fragment is dominant (b) so annihilates the $x$ fragment as they collide (c) then spreads across the north and east channels (d) then breaks into 2 fragments which travel along the north and east channels (grey arrows show the direction of the fragment input).}
\label{fig20}
\end{figure}

Figure~\ref{fig19}a--d shows the results for $\langle x,y \rangle = \langle 1,1 \rangle$ at a light intensity of 
1.5mW cm$^{-2}$ in the excitable domain. In this implementation the fragment $x$ is dominant (despite the $y$ 
input channel being shorter) and it undergoes collision with fragment $y$, resulting in annihilation of the $y$ fragment. However, due to the excitability of the channels the fragment expands in a north east direction at the junction and splits into two fragments that travel along the output $p$ and $q$ channels. Figure~\ref{fig20}a--d shows the converse case where the $y$ fragment is dominant and annihilation of the $x$ fragment occurs and the resulting daughter fragment also travels towards north east direction then breaks into 2 fragments which travel along the output $p$ and $q$ channels at light intensity 3.0mW cm$^{-2}$ (channels are weakly excitable).

\begin{figure}[!tbp] 
\centering
	\subfigure[]{
		\includegraphics[height=35mm]{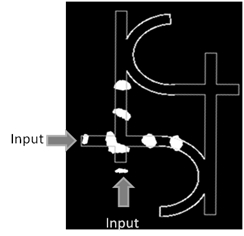}	
	}
	\subfigure[]{
		\includegraphics[height=35mm]{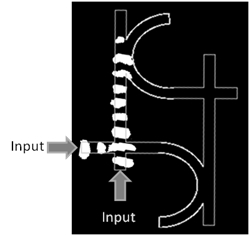}	
	}
\caption{When $x=1$ and $y=1$ at light intensity 3.1mW cm$^-$$^2$ (a) both fragments collide to give a V-shaped fragment travelling in a north east direction, progression of fragment every 30 seconds (b) $x$ fragment is annihilated by the $y$ fragment and the resulting daughter fragment travels north into the $q$ output channel only, progression of fragment every 20 seconds.}
\label{fig21}
\end{figure}

Figure~\ref{fig21}a shows the case when both input fragments collide at the same time at the cross junction, collide to form a daughter fragment which splits into two fragments at the cross junction and travels along the output $p$ and $q$ channels even when the excitability is relatively low (light intensity 3.1mW cm$^{-2}$) (Fig.~\ref{fig21}a). However, at the same light level if the $y$ fragment is slightly dominant then the $x$ fragment is annihilated and the resulting daughter fragment travels into the output $q$ channel only, see Fig.~\ref{fig21}. This is the correct implementation for the half adder scheme to be solved. However, this emphasises the importance of synchronising the timing of the two inputs in the experiments. This is difficult relative to the simulation experiments although differences in the position of the initiations relative to the input channels gave similar results in the simulation experiments. In the experiment we utilise a constant fragment generator in a dark region which is then linked to two same length excitable channels that transfer fragments to the inputs. These channels are removable and in theory should deliver one fragment to each input at the same time interval. However, even a small difference in timing has a marked effect on the operation of the cross-junction as the lengths of the $x$ and $y$ input channels are supposed to confer the functionality of the gate. This error in timing is compounded by the fact that the input channels are only a few mm in length. The results do show that a wide range of behaviour is possible for the same junction by altering the light level and input timing. At a critical light level and with the correct timing (i.e. both fragments enter their respective input channels at the same time) then the cross junction is successfully implemented according to the half adder scheme (where they must provide the appropriate inputs for gates P1).

\subsection{Attempt to construct one-bit half-adder in experiment}

\begin{figure}[!tbp] 
\centering
	\subfigure[]{
		\includegraphics[height=35mm]{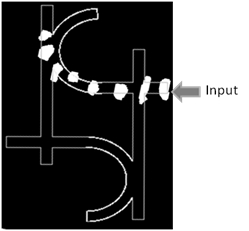}	
	}
	\subfigure[]{
		\includegraphics[height=35mm]{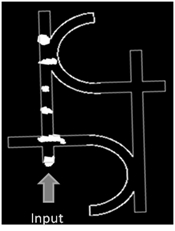}	
	}
	\subfigure[]{
		\includegraphics[height=35mm]{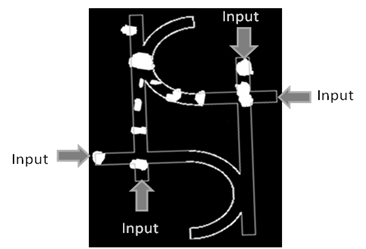}	
	}
\caption{Progression of fragments seen every 30 seconds at sub excitable level when (a) $x=1$ (with one $x$ input only) at light intensity 3.1mW cm$^-$$^2$ and (b) $y=1$ (with one $y$ input only) at light intensity 3.1mW cm$^-$$^2$, 
(c) $x=1$ and $y=1$ (with fragments at both $x$ inputs and both $y$ inputs simultaneously) at light intensity 3.1mW cm$^-$$^2$.}
\label{fig22}
\end{figure}

Figure~\ref{fig22} shows the progression of fragments seen every 30~sec at a sub-excitable level, light intensity 
3.1mW cm$^{-2}$ when (a)~$\langle x,y \rangle = \langle 1,0 \rangle$ with one $x$ input only and 
(b)~$\langle x,y \rangle = \langle 0,1 \rangle$ with one $y$ input only, (c)~$\langle x,y \rangle = \langle 1,1 \rangle$ with fragments at both $x$ inputs and both $y$ inputs simultaneously. As previously observed at this light level when solving the interaction ates P1 and P2 an input in either channel $x$ or $y$ results in an output in either channel $p$ or $q$ respectively thus giving the correct output according to this theoretical construct of a one-bit half-adder. Although only one $x$ and $y$ input was initiated for ease of experimental implementation each part of the junction has been solved previously so solving two simultaneous $x$ or $y$ inputs in experiment at this light level should be achievable. For the case where $x=1$ and $y = 1$ both sets of inputs were initiated simultaneously in order to implement the half bit adder. Both respective cross junctions were solved successfully in this experimental implementation. However, despite a collision prior to the two output channels a fragment could only be observed in one output channel $q$. Therefore, the entire half bit adder scheme is not solved at a single light level in experiment.  If we revisit the fractional gate implementation in experiment we see that this is actually in agreement with the results from Fig.~\ref{fig14} where the collision sequence where $x=1$ and $y =1$ is only solved at higher excitability levels. Therefore, the half bit adder could be solved in experiment via projection of differential light levels (or by altering the light level online) in order to achieve all parts of the adder scheme in combination. If we compare this to the simulation results each part of the adder $x=1$, $y=1$ and $x=1$ and $y=1$ are solved at subtly different values of $\phi$ all close to the sub-excitable limit. In experiment we don't have such fine control over the light dependent factor meaning that even if a single light level to solve the adder scheme when x and y = 1 does exist (as predicted by simulation) the degrees of freedom in altering the light level make finding it improbable especially when considering the heterogeneity of the chemical system (which has limited controllability).                                     
                                    
\section{Summary and discussion}
\label{discussion}

We demonstrated in numerical simulations that the light-sensitive BZ medium near a sub-excitable level realizes a wide variety of logical operations when weakly-constrained by illumination-induced geometrical architectures.  We designed two types of Boolean logic gates, both gates have two inputs and two outputs. The gates implement transformations  $\langle  x, y \rangle \rightarrow \langle  xy, x+y \rangle$ and  $\langle  x, y \rangle \rightarrow \langle  x, \overline{x}y \rangle$. In addition by mapping all possible light levels and input states of these interaction gates a number of intermediate output states were identified showing that complex ``light mediated'' logic gates can be constructed from relatively simple interaction gates. We shown how the interaction gates can be assembled into a
one-bit half-adder. The full functionality of the adder is illustrated using a two-variable Oregonator model modified to account for light sensitivity. The different input states of the adder were implemented using different values of$\phi$. This fine control of $\phi$ prevents wave-fragments from expanding in the propagation channels and junction-chambers in order to confer the correct functionality of the integral parts of the adder. This is the first instance of arithmetic circuits being implemented in models of excitable chemical systems. 
In laboratory experiments we were able to implement the interaction gates that make up the one-bit half-adder scheme. We were able to show that in agreement with simulation results the functionality of the simple gate is sensitively dependent on the light level. Therefore, for the same input states we were able to show various output states dependent on the light level.  Even though some of the functions of the interaction gates differed between experiment and simulation they were in close agreement when considering the functionality of the interaction gate needed to solve the adder effectively. Therefore, in both the simulation and experimental implementations a high level of illumination close to the sub-excitable limit produced a one input/one output state for both gates P1 and P2. In addition they produced a two input/one output state for gate p2 provided the input timing was correct. For gate P1 the correct implementation was two input/two outputs but this was only achieved at relatively low light levels in experiment compared to simulation experiments. In laboratory experiments to implement the one-bit half-adder there was not a single light level that gave the correct implementation. In some respects this agrees with simulation results as for different input sequences different levels of $\phi$ are required to stabilise the wave fragments. Therefore, there is ot a single value of $\phi$ which gives the correct outputs for all possible input sequences. However, in the simulation all the light levels used are close to the sub-excitable level whereas in experiment to implement the final collision sequence in gate P1 requires higher levels of excitability (low light levels). Therefore, for the interaction gates P1 and P2 to work in combination the light level would have to be projected differentially or changed online in order to implement the adder correctly. Therefore, work will continue to identify experimental parameters or different gate schemes whereby arithmetic circuits can be implemented effectively in experiment. These laboratory experiments have shown that the control of propagating wave-fragments is extremely difficult due to the fragments very high sensitivity to environmental conditions. In this case the difficulties were multiplied as solving the adder scheme required precision timing of the inputs and the sub-excitable nature of the chemical media meant that the fragment stability was so low that the whole scheme had to be solved at a mm scale. Therefore our future task will be to increase the stability of the wave-fragments and to realise correct functioning arithmetic circuits in experiment and many-bit full adders in simulation and experiment.


\begin{thebibliography}{99}


\bibitem{cbc}
Adamatzky~A. (Ed.) Collision-Based Computing. 
Springer, 2003.

	\bibitem{adamatzky_2004_collision} Adamatzky A. Collision-based
  computing in Belousov--Zhabotinsky medium. Chaos Solitons Fractals
  21 (2004) 1259--1264

\bibitem{RDCBook}
Adamatzky~A., De~Lacy~Costello~B., Asai~T. 
Reaction-Diffusion Computers (Elsevier, 2005).

	\bibitem{andy_ben_BZ_collision} Adamatzky A., and De Lacy Costello B.
  Binary collisions between wave-fragments in a sub-excitable
  Belousov-Zhabotinsky medium. Chaos, Solitons \& Fractals 34 (2007)
  307--315.

\bibitem{adamatzkyPrivmanIssue}
Adamatzky~A. 
Topics in reaction-diffusion computers.
J Comput Theor NanoSciences (2010), in press.


\bibitem{adamatzkyKazIssue}
Adamatzky~A., De Lacy Costello~B., Bull~L., Holley~J.
Towards arithmetic circuits in sub-excitable chemical media
Israel J Chemistry (2010), in press.


\bibitem{adamatzkyPhysGate}
Adamatzky~A. Slime mould logical gates: exploring ballistic approach (2010).
\url{arXiv:1005.2301v1 [nlin.PS]}

\bibitem{adamatzkyPhysarumMachines}
Adamatzky~A. Physarum Machines (World Scientific, 2010). 


	\bibitem{beato_engel} Beato~V., Engel~H. Pulse propagation in a model
  for the photosensitive Belousov-Zhabotinsky reaction with external
  noise. In: Noise in Complex Systems and Stochastic Dynamics, Edited
  by Schimansky-Geier~L., Abbott~D., Neiman~A.,
  Van~den~Broeck~C. Proc. SPIE 5114 (2003) 353--362.
  
    \bibitem{berlekamp_1992}
	Berlekamp E.R., Conway J.H., Guy R.L.
	Winning ways for your mathematical plays, vol. 2. Academic Press; 1982.
  
  \bibitem{ben_gun} De~Lacy~Costello~B., Toth~R., Stone~C.,
  Adamatzky~A., Bull~L.  Implementation of glider guns in the
  light-sensitive Belousov-Zhabotinsky medium Phys. Rev. E 79 (2009)
  026114.
  

	\bibitem{field_noyes_1974} Field~R.~J., Noyes~R.~M.  Oscillations in
  chemical systems. IV. Limit cycle behavior in a model of a real
  chemical reaction.  J. Chem. Phys. 1974 (60) 1877--1884.
  
  \bibitem{fredkin_toffoli_1982}
  Fredkin F, Toffoli T. Conservative logic. 
  Int J Theor Phys 21 (1982) 219-–253.
  

 \bibitem{gorecka_2003} G\'{o}recka J. N., G\'{o}recki J.  T-shaped
  coincidence detector as a band filter of chemical signal frequency,
  Phys. Rev. E 67 (2003) 067203.

\bibitem{gorecki_2003} G\'{o}recki J., Yoshikawa K. and Igarashi Y.,
  On chemical reactors that can count, J. Phys. Chem. A 107 (2003)
  1664--1669.

\bibitem{gorecki_2005} G\'{o}recki J., G\'{o}recka J. N., Yoshikawa
  K., Igarashi Y., Nagahara H.  Sensing the distance to a source of
  periodic oscillations in a nonlinear chemical medium with the output
  information coded in frequency of excitation pulses. Phys. Rev. E 72
  (2005) 046201.

\bibitem{gorecki_2006} G\'{o}recki J. and G\'{o}recka J. N.,
  Multi-argument logical operations performed with excitable chemical
  medium, J. Chem. Phys. 124 (2006) 084101.

\bibitem{gorecki_2006a} G\'{o}recki J., G\'{o}recka J. N.  Information
  processing with chemical excitations --- from instant machines to an
  artificial chemical brain Int J Unconv Comput 2 (2006) 321--336.

\bibitem{gorecki_2009} G\'{o}recki~J., G\'{o}recka~J.~N., Igarashi~Y.
  Information processing with structured excitable medium, Natural
  Computing 8 (2009) 473--492.

\bibitem{gorecka_2007} G\'{o}recka J. N., G\'{o}recki J., Igarashi Y.
  On the simplest chemical signal diodes constructed with an excitable
  medium, Int J Unconventional Computing 5 (2009) 129--143.

\bibitem{margolus}
Margolus N. Physics-like models of computation. Physica D 10 (1984) 81–-95.


  \bibitem{motoike_2003} Motoike~I.~N. and Yoshikawa~K.  Information
  operations with multiple pulses on an excitable field.  Chaos,
  Solitons \& Fractals 17 (2003) 455--461.
  
   \bibitem{sendina_2001} 
Sendi\H{n}a-Nadal~I., Mihaliuk~E.,
Wang~J., P\'{e}rez-Mu\H{n}uzuri~V. and Showalter~K. Wave
propagation in subexcitable media with periodically modulated
excitability. Phys. Rev. Lett. 86 (2001) 1646--1649.

\bibitem{sielewiesiuk_2001} Sielewiesiuk J. and G\'{o}recki J.,
  Logical functions of a cross junction of excitable chemical media,
  J. Phys. Chem., A105 (2001) 8189.

\bibitem{toth1994} T\'oth A., G\'asp\'ar V. and Showalter K.
Propagation of chemical waves through capillary tubes. J. Phys.
Chem. 98 (1994) 522--531.
  
  
\bibitem{toth_showalter} T\'{o}th A. and Showalter K. Logic gates
in excitable media. J. Chem. Phys. 103 (1995) 2058--2066.


\bibitem{RITABEN} Toth~R., Stone~C., Adamatzky~A., de Lacy
  Costello~B., Bull~L.  Experimental validation of binary collisions
  between wave-fragments in the photosensitive Belousov-Zhabotinsky
  reaction.  Chaos, Solitons \& Fractals 41 (2009) 1605--1615.

\bibitem{yoshikawa_2009} Yoshikawa~K., Motoike~I.~M., Ichino~T.,
  T. Yamaguchi, Y. Igarashi, J. Gorecki and J. N. Gorecka Basic
  information processing operations with pulses of excitation in a
  reaction-diffusion system.  Int J Unconventional Computing 5 (2009)
  3--37.
 
\bibitem{yoshikawa_2009a} Yoshikawa~K., Nagahara~H., Ichino~T.,
  J. Gorecki, J. N. Gorecka and Y. Igarashi On chemical methods of
  direction and distance sensing.  Int J Unconventional Computing 5
  (2009) 53--65.


\end{thebibliography}
\end{document}